\documentclass[]{article}

\usepackage{graphicx}
\usepackage{amsmath}
\usepackage{amssymb}
\usepackage{times}
\usepackage{braket}

\title{Prospects for measuring the electric dipole moment of the electron using electrically trapped polar molecules}

\author{M. R. Tarbutt, J. J. Hudson, B. E. Sauer and E. A. Hinds\\[3mm]
Centre for Cold Matter, Blackett Laboratory,
Imperial College London,\\Prince Consort Road,
London SW7 2AZ, United Kingdom.}

\date{}

\begin{document}

\maketitle

\renewcommand{\thefootnote}{\fnsymbol{footnote}}

\line(1,0){340}\\

\noindent Heavy polar molecules can be used to measure the electric dipole moment of the electron, which is a sensitive probe of physics beyond the Standard Model. The value is determined by measuring the precession of the molecule's spin in a plane perpendicular to an applied electric field. The longer this precession evolves coherently, the higher the precision of the measurement. For molecules in a trap, this coherence time could be very long indeed. We evaluate the sensitivity of an experiment where neutral molecules are trapped electrically, and compare this to an equivalent measurement in a molecular beam. We consider the use of a Stark decelerator to load the trap from a supersonic source, and calculate the deceleration efficiency for YbF molecules in both strong-field seeking and weak-field seeking states. With a 1\,s holding time in the trap, the statistical sensitivity could be ten times higher than it is in the beam experiment, and this could improve by a further factor of five if the trap can be loaded from a source of larger emittance. We study some effects due to field inhomogeneity in the trap and find that rotation of the electric field direction, leading to an inhomogeneous geometric phase shift, is the primary obstacle to a sensitive trap-based measurement.

\line(1,0){340}

\section{Introduction}

The permanent electric dipole moment (edm) of the electron, or other fundamental particle, can only be non-zero if both parity (P) and time-reversal (T) invariance are violated \cite{Purcell(1)50,Landau(1)57}. The weak interaction violates P, while T violation is equivalent to CP violation provided CPT invariance is accepted. The CP violation of the Standard Model generates edms that are far too small to detect \cite{Pospelov(1)91}. It also fails to account for the predominance of matter over antimatter in the universe \cite{Barr(1)79}. Many of the proposed extensions of the Standard Model contain new sources of CP violation and result in vastly larger edms \cite{Bernreuther(1)91, Commins(1)99}; indeed, some models predict values for the electron edm that are very close to the current experimental limit. A measurement of a non-zero edm would be firm evidence for new physics.

Heavy, paramagnetic atoms and molecules offer high sensitivity to the electron edm. In an applied electric field, $E_{a}$, the change in energy of such an atom or molecule resulting from the electron edm, $d_{e}$, is $d_{e}P(E_{a}) E_{\rm int}$. Here, $E_{\rm int}$ is a structure-dependent effective electric field whose magnitude scales as the cube of the nuclear charge \cite{Hinds(1)97}, and $P(E_{a})$ is the degree of polarization of the atom or molecule in the applied field. The most sensitive measurement to date, $d_{e}=(6.9 \pm 7.4)\times 10^{-28}e$\,cm, was made using a beam of Tl atoms \cite{Regan(1)02}, for which $E_{\rm int}$ is large but $P$ is small for realistic laboratory fields. Because polar molecules can have much larger values of $P$ they offer even higher sensitivity \cite{Sandars(1)67, Sandars(1)75,Sushkov(1)78}. Using a continuous, thermal beam of YbF, we made the first measurement of the electron edm with a molecular system \cite{Hudson(1)02}. By upgrading this experiment to a cold, pulsed source, its statistical sensitivity has been greatly improved, and a new measurement is underway \cite{Hudson(1)05,Sauer(1)06}. Using the most recent advances in the production and manipulation of cold molecules, it seems likely that still higher precision can be obtained. The ability to trap polar molecules electrically \cite{Bethlem(2)00} is particularly attractive for precision measurements such as the electron edm, because of the very long interaction times that are then available.

The shot-noise limit on the statistical error in a molecule-based measurement of the edm, $\delta d_{e}$, in the usual units of $e$\,cm, is given by
\begin{equation}
\delta d_{e}=\frac{\hbar}{e} \frac{1}{|P| E_{\rm{int}} \tau \sqrt{N}}.
\label{Eq:SensitivityFormula}
\end{equation}
Here, $E_{\rm{int}}$ is measured in V/cm, $\tau$ is the amount of time each molecule spends inside the apparatus (the coherence time), and $N$ is the total number of molecules that participate in the experiment. The inefficiencies of state preparation and readout, and any intervening state manipulations, are incorporated in our definition of $N$. In our current edm experiment, pulses of YbF molecules created at 4\,K in a supersonic source travel at 600\,m/s through a 60\,cm long interaction region where the applied electric field is 20\,kV/cm. There are 25 pulses per second, approximately 2500 molecules are detected in each pulse, and 50\% of the running time is devoted to data-taking. For YbF, $E_{\rm{int}}=26$\,GV/cm, and the polarization factor at 20\,kV/cm is $|P|=0.7$. Using these values in Eq.\,(\ref{Eq:SensitivityFormula}), we estimate our shot-noise limited sensitivity to be $7\times 10^{-28}\,e\,\rm{cm}/\sqrt{\rm{day}}$. As is often the case, the experiment operates slightly above the shot noise limit, magnetic field noise and source intensity fluctuations being the dominant sources of additional noise. At present, our statistical error is about 1.3 times the shot-noise limit.

To increase the sensitivity, we have to increase the product $\sqrt{N}\tau$. There are, of course, many possible ways to do this. An increase in $N$ might be obtained by improving the detection efficiency or by increasing the time-averaged flux of molecules from the source. These would benefit both the beam experiment and the trap experiments considered here. Another strategy is to increase the fraction of molecules transmitted from source to detector, which might be achieved using molecular optics near the source to collimate the beam. For that to yield an improvement, the transverse phase space acceptance of the optics must exceed that of the machine without the optics. We will discuss beam optics below in the context of Stark decelerators, and we will see that the acceptance of our current machine is already comparable to that of realistic beam optics. An increase in $\tau$ could be achieved without slowing or trapping the molecules, simply by making the machine longer. For a freely diverging molecular beam however, there is no advantage to increasing the interaction length $L$ since the coherence time is proportional to $L$ and the number of molecules reaching the detector scales as $L^{-2}$, leaving $\tau\sqrt{N}$ constant. If optics are used near the source to collimate or focus the molecular beam, then an increase in $L$ can be beneficial.

In this paper, motivated by the prospect of an enormous increase in coherence time, we consider the possibility of measuring the electron edm using electrically trapped molecules. Given the difficulties involved in producing trappable molecules, we can expect the increase in $\tau$ to be accompanied by a reduction in $N$. We evaluate the efficiency of Stark deceleration and so obtain an estimate for the statistical sensitivity that could be obtained. We then consider a few of the very many systematic effects that may scupper the measurement. We focus mainly on a measurement using YbF molecules, though our considerations are easily modified to other relevant molecular systems.

\section{Deceleration}

If the molecules are to be trapped, they must have a low velocity. We consider the prospects of bringing to rest the YbF molecules produced by our current supersonic source \cite{Tarbutt(1)02}. We take 340\,m/s and 4\,K as the starting speed and temperature, these being typical when using room temperature xenon as the carrier gas. We have obtained speeds as low as 290\,m/s by cooling the xenon, but only at the expense of molecular flux. It seems possible that a buffer gas source \cite{Maxwell(1)05,Patterson(1)07} will provide slower YbF beams with higher intensity, but this has not yet been demonstrated.

\begin{figure}[t]
\begin{center}
\includegraphics[width=12cm]{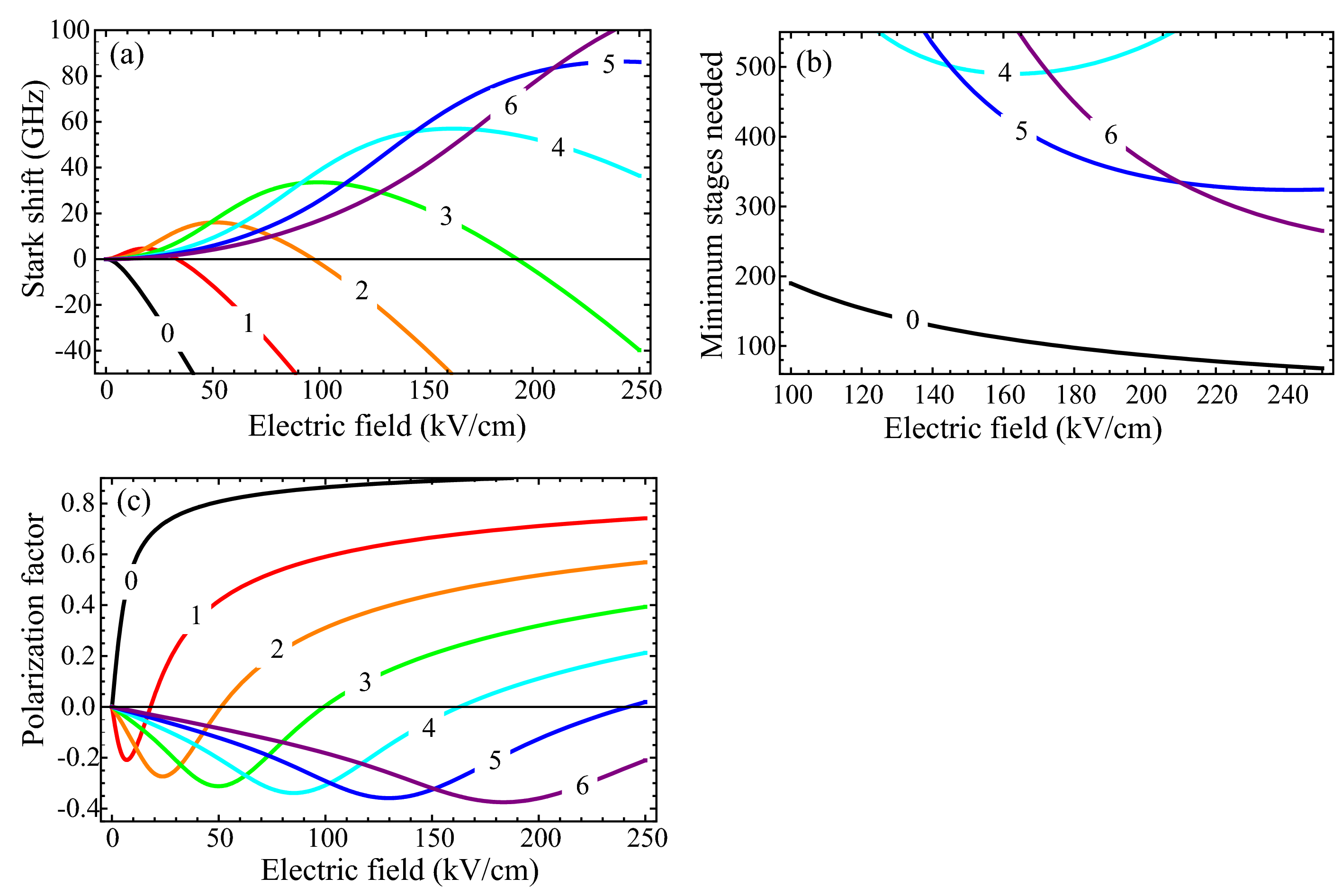}
\caption{
\label{Fig:Stark}
Electric field dependence of (a) the Stark shifts of $(N\!=\!0\!-\!6,M\!=\!0)$ states of YbF, (b) the minimum number of deceleration stages needed to stop 340\,m/s YbF molecules in a selection of these states, and (c) the polarization factor of each state. Each curve is labelled by the value of the rotational quantum number, $N$.}
\end{center}
\end{figure}

In designing a suitable decelerator, we can choose which rotational state of the molecule to use. Figure \ref{Fig:Stark}(a) shows the Stark shifts of the $(N,M\!=\!0)$ rotational states of YbF in fields up to 250\,kV/cm. Here, $N$ and $M$ are the quantum numbers of the rotational angular momentum and its projection onto the electric field axis. In the ground state the molecule has a negative Stark shift and can only be decelerated using the alternating gradient (AG) method \cite{Bethlem(1)02,Tarbutt(1)04,Bethlem(1)06}. The other states shown have positive Stark shifts at low field, and negative Stark shifts at high field. Molecules prepared in these excited rotational states could be slowed down with a ``conventional'' Stark decelerator for weak-field seekers \cite{Bethlem(1)99}, which we will refer to as a WF decelerator. It would have to be operated at fields below the turning point of the state. Figure \ref{Fig:Stark}(b) shows the minimum number of deceleration stages needed to decelerate YbF molecules from 340m/s to zero. Deceleration in the ground state (Sec.\,\ref{Sec:AG}) is obviously advantageous in terms of stages needed, but the difficulty of implementing the AG method is a disadvantage. Deceleration in the weak-field seeking states (Sec.\,\ref{Sec:WF}) is easier to implement but requires more stages. The best choice of state depends on the operating field. At 200\,kV/cm\footnote{We have operated decelerators at this field. Higher fields are very difficult to sustain.} the number of WF deceleration stages is minimized for $N\!=\!5$. Lower rotational states are not as efficient because their turning points are reached at lower fields, as illustrated by the curve for $N=4$ in the figure. Higher rotational states are also less efficient because their Stark shifts are smaller at this field, as illustrated by the curve for $N=6$.

We have looked at the rotational state dependence of deceleration, but what about the edm experiment itself? Its sensitivity depends on the rotational state and the applied electric field through the polarization factor. This is given in terms of the Stark shift, $W$, the electric field, $E$, and the dipole moment, $\mu$, by $P=-(1/\mu)\,dW/dE$. Figure \ref{Fig:Stark}(c) shows this polarization factor as a function of electric field for the various rotational states with $M\!=\!0$. At any given field the ground state offers the largest polarization factor, which is one reason why our current experiment uses this state. An experiment that used a weak-field seeking state would have a smaller, though still useful, value of $|P|$. For example, the $N=5$ state discussed above has $|P|=0.36$ at $E=130$\,kV/cm.

\subsection{Deceleration in a rotationally excited state}\label{Sec:WF}

In this section we consider a practical decelerator for molecules in the $(N,M)=(5,0)$ state. We consider a design with a 4\,mm square aperture and a periodicity of $2L=24$\,mm, similar to the one used in Berlin to decelerate OH \cite{Meerakker(1)05}. Using voltages of $\pm40$\,kV, and a synchronous phase angle of $66^{\circ}$, 444 stages are required to decelerate from 340\,m/s to zero. Each unit cell of the decelerator has two deceleration stages, one that focusses along $x$, and the other along $y$, so the overall length is 5.328\,m. These are the parameters we use in the rest of this section.

To estimate how many molecules we can expect to decelerate, we need to calculate the phase-space acceptance of the decelerator. We start with the longitudinal acceptance. Since for our case the Stark shift is not at all linear (see Fig.\,\ref{Fig:Stark}), we will need to take a short diversion to extend the usual treatment of the longitudinal acceptance e.g. \cite{Bethlem(1)00,Meerakker(2)05}. The change in kinetic energy of the synchronous molecule in each deceleration stage is usually well approximated by $\Delta K = -\Delta K_{\rm{max}} \sin \phi_{0}$, with $\Delta K_{\rm{max}}$ being the maximum possible energy change per stage (a positive number), and $\phi_{0}$ the synchronous phase angle which goes from 0 to $2\pi$ over the period $2L$. For our case, this is a poor approximation, and we find it necessary to retain an extra term in the Fourier expansion of $\Delta K$. We shall derive a general formula for the longitudinal acceptance, and then apply it to our case. We write the change in kinetic energy using the expansion
\begin{equation}
\Delta K = \Delta K_{\rm{max}}\sum_{n\text{ odd}}(-1)^{(n+1)/2}a_{n} \sin(n \phi_{0}).
\label{Eq:DeltaK}
\end{equation}
Here, the $a_{n}$ are coefficients that depend on the details of the decelerator and Stark shift of the molecule, and satisfy $\sum_{n\text{ odd}}a_{n}=1$. Next, we use the approximation of a constantly acting force to write down the equation of motion of a non-synchronous molecule in terms of its relative phase, $\tilde\phi$, and its relative velocity $\tilde v$,
\begin{equation}
m \pi \tilde v \frac{d \tilde v}{d \tilde \phi}=\Delta K_{\rm{max}}\sum_{n\text{ odd}}(-1)^{(n+1)/2}a_{n}\left[\sin(n \phi_{0}+ n \tilde\phi)-\sin(n\phi_{0})\right],
\end{equation}
where $m$ is the mass. Integrating this equation of motion we obtain the contours of constant (relative) energy,
\begin{equation}
\left(\frac{\tilde v}{v_{i}}\right)^{2}+\frac{1}{\pi N_{\min}}\sum_{n\text{ odd}}(-1)^{(n+1)/2}a_{n}\left[\frac{1}{n}\cos(n \phi_{0}+ n \tilde\phi)+\tilde\phi\sin(n\phi_{0})\right]=\rm{constant},
\label{Eq:RelativeEnergy}
\end{equation}
where $v_{i}$ is the initial velocity and $N_{\min}$ is the minimum number of deceleration stages needed to bring the molecules to rest. The potential well represented by the second term in Eq.\,(\ref{Eq:RelativeEnergy}) has a maximum at $\tilde\phi = \pi - 2\phi_{0}$, irrespective of the values of the $a_{n}$, for any sensible decelerator configuration. This is easily verified by inspecting the derivative of this potential with respect to $\tilde\phi$. A molecule that reaches this maximum with zero relative velocity is moving on the separatrix between bound and unbound motion. The equation of the separatrix is thus found to be
\begin{multline}
\left(\frac{\tilde v}{v_{i}}\right)^{2}+\frac{1}{\pi N_{\min}}\sum_{n\text{ odd}}(-1)^{(n+1)/2}a_{n}
\left[\frac{2}{n}\cos\left(n \phi_{0} + \tfrac{1}{2}n \tilde\phi\right)\cos\left(\tfrac{1}{2}n\tilde\phi\right)\right.\\
\left.+(2\phi_{0}+\tilde\phi-\pi)\sin(n\phi_{0})\right]=0.
\label{Eq:separatrix}
\end{multline}
\begin{figure}[t]
\begin{center}
\includegraphics[width=12cm]{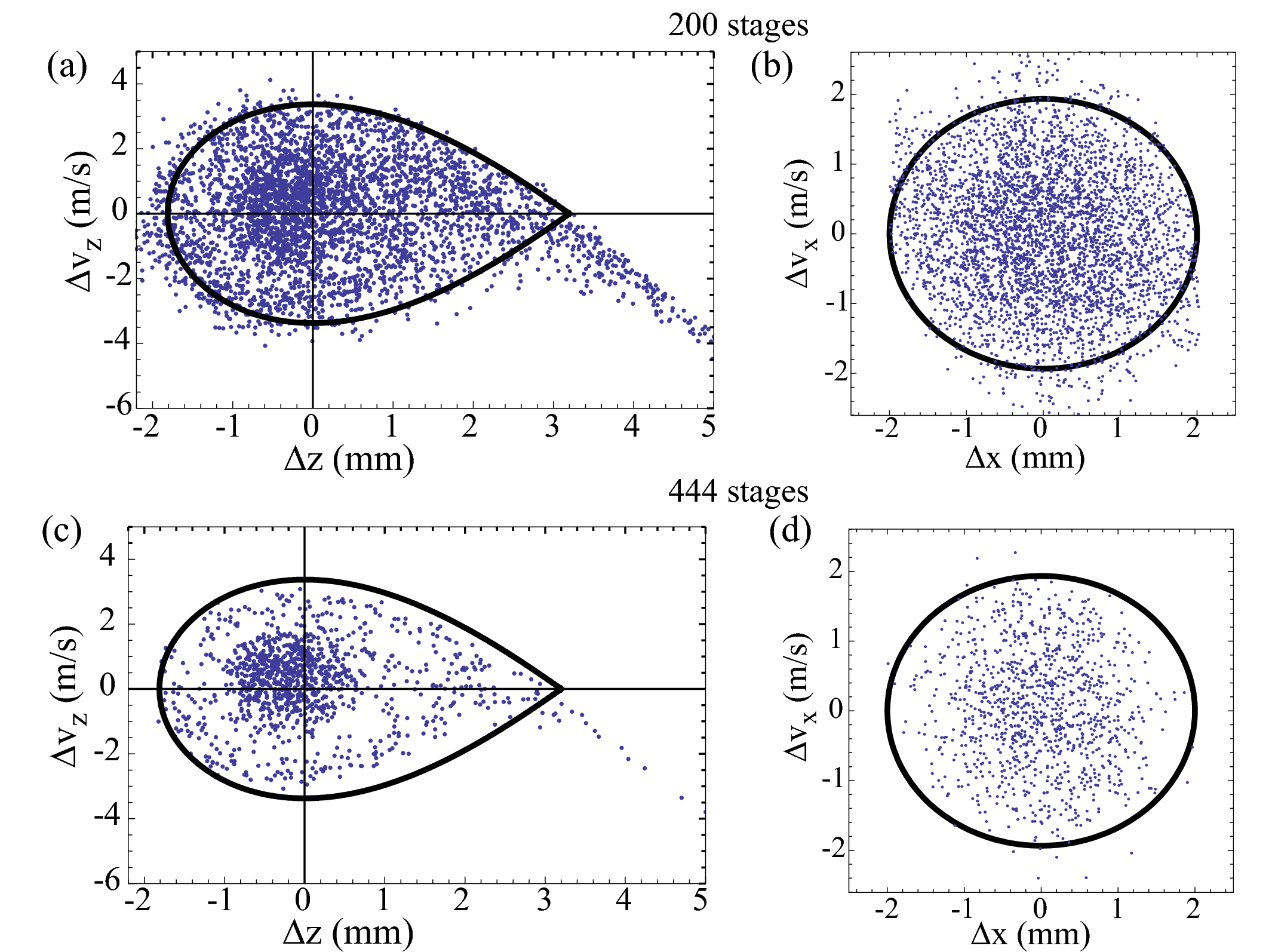}
\caption{
\label{Fig:WFAccept}
Phase-space acceptance of a Stark decelerator for YbF in the $(5,0)$ state. Dots show the initial positions in phase space of those molecules transmitted and decelerated, as determined from a three dimensional numerical simulation. Bold lines show the acceptances calculated analytically as discussed in the text. (a, c) Longitudinal acceptance after 200, 444 stages. (b, d) Transverse acceptance after 200, 444 stages.}
\end{center}
\end{figure}

For our case, the first two terms in this sum give a sufficiently accurate result. Using the electric fields obtained from a finite element model, and the Stark shift shown in Fig.\,\ref{Fig:Stark}(a), we can determine $\Delta K$ numerically. Fitting this to the first two terms of Eq.\,(\ref{Eq:DeltaK}), with $a_{3}=1-a_{1}$, we obtain $a_{1}=0.825$. The bold line in Fig.\,\ref{Fig:WFAccept}(a,c) shows the longitudinal acceptance area calculated using the first two terms of Eq.\,(\ref{Eq:separatrix}), and this value of $a_{1}$. Integrating the area inside this separatrix gives a longitudinal phase-space acceptance of $A_{z}=23.4$\,mm\,m/s. Note that the calculated acceptance area is smaller, $15.2$\,mm\,m/s, when only the first term in Eq.\,(\ref{Eq:DeltaK}) is used (i.e. $a_{1}=1$).

Next, we estimate the transverse acceptance, following the procedure outlined in e.g. \cite{Bethlem(1)02,Meerakker(1)06}. The restoring force in the transverse direction is approximately linear in the transverse coordinate, $F_{x}\simeq -k_{x}(z)x$, but with a spring constant that is a strongly-varying function of the longitudinal coordinate, with period $2L$. The wavelength of the transverse oscillation is very much longer than $2L$, except when the molecules have been decelerated to very low speeds. We calculate a mean spring constant, $\bar k_{x}$, by integrating $k_{x}(\phi)$ from $\phi_{0}-\pi$ to $\phi_{0}$ in the first switch state, and then from $\phi_{0}$ to $\phi_{0}+\pi$ in the second switch state, finally dividing by $2\pi$. In this way, we find the mean transverse angular oscillation frequency for our molecules of mass $m$ to be $\omega_{x}=\sqrt{\bar k_{x}/m}=2\pi \times 154$\,Hz when $\phi_{0}=66^{\circ}$. The corresponding phase-space acceptance in the $x$-direction is $A_{x}=\pi\omega_{x} r_{0}^{2}$, where $r_{0}=2$\,mm is half the size of the decelerator's aperture. Thus, this approximate calculation gives us $A_{x}=12.2$\,mm\,m/s. The bold line in Fig.\,\ref{Fig:WFAccept}(b,d) shows the transverse acceptance obtained using this approximate method. The total 6D acceptance is $A=A_{x}A_{y}A_{z}=A_{x}^{2}A_{z}=3487\,\rm{mm}^{3}\,(\rm{m/s})^{3}$.

The acceptance of a real Stark decelerator tends to be less than the idealized estimates given above. Coupling of the longitudinal and transverse motions can lead to unstable regions in phase-space \cite{Meerakker(1)06}. Further loss occurs at very low speeds as the wavelength of transverse oscillations becomes comparable to the decelerator periodicity and the molecules are thrown out of the decelerator in regions where the focussing is weak \cite{Sawyer(1)08}. To calculate the true phase-space acceptance, we simulated the motion of molecules through the decelerator. Figure \ref{Fig:WFAccept} shows the results obtained for 200 stages, where the final speed is 252\,m/s, and for 444 stages where the final speed is 14\,m/s, slow enough that one final stage - the trap - will bring the molecules to rest. To produce these plots, the trajectories of molecules chosen at random from a larger phase-space volume were simulated through the decelerator. The initial coordinates of successful trajectories, those that reached the end {\it and} were decelerated, are marked by the dots. The acceptance is the volume of the initial distribution multiplied by the successful fraction. After 200 stages, the region of phase space that contributes to the slowed beam is very similar to that predicted by the simple theory outlined above. In the longitudinal direction, part (a), some molecules are accepted from regions that are slightly outside the bold line. This is because off-axis molecules that travel closer to the rods of the decelerator experience a slightly deeper longitudinal potential than those on the axis. The region inside the bold line is not uniformly filled, the area closer to the origin having a higher density because it is less susceptible to coupling of the transverse and longitudinal motions. In the transverse direction, part (b), we see again that a few molecules outside the region given by the simple theory contribute to the acceptance because the transverse confinement is greater for some of the non-synchronous molecules. Once again, the central region has a higher density of accepted particles. The total 6D phase-space acceptance for 200 stages is 1168\,$\rm{mm}^{3}\,(\rm{m/s})^{3}$, which is a factor of 3 smaller than the idealized prediction. An increase in the number of stages to 444, so that the molecules can be trapped, reduces the acceptance to 352\,$\rm{mm}^{3}\,(\rm{m/s})^{3}$. In the longitudinal direction, Fig.\,\ref{Fig:WFAccept}(c), the coupling of transverse and longitudinal motions results in an unstable region of phase-space within the separatrix, as investigated in \cite{Meerakker(1)06}. The accepted molecules come from a region close to the origin, along with the `halo' region around this. In the transverse direction, part (d), molecules coming from regions close to the rods of the decelerator tend to be focussed too strongly at low speed and so are lost.

In our supersonic beam, the rotational temperature is typically about 4\,K, and the population in the $(5,0)$ state is only a few percent of the ground state population. To increase the flux of slow molecules, we will need to transfer population into our chosen state. This could be done by driving the microwave transition from the ground state in the presence of a static electric field. Without the static field it is impossible to drive this transition, since the total angular momentum quantum number has to change by 5 units. In the presence of a field, states of the same $M$ but different $N$ are mixed and the selection rule on $N$ is no longer a good one. In the limit of very strong fields the molecules are strongly polarized along the field axis and the eignestates, known as pendular states, are those of a two-dimensional angular oscillator \cite{Rost(1)92}. They are labelled by $M$ and by a vibrational quantum number $v_{p}$, which only changes by one unit in an electric dipole transition. The state that correlates to the field-free $(5,0)$ state is the one with $(v_{p},M)=(10,0)$, while the ground state has $(v_{p},M)=(0,0)$. It follows that the required transition is forbidden in the strong-field limit as well as in the weak-field limit. At intermediate fields, it may be possible to drive the transition.

\begin{figure}
\begin{center}
\includegraphics[width=8cm]{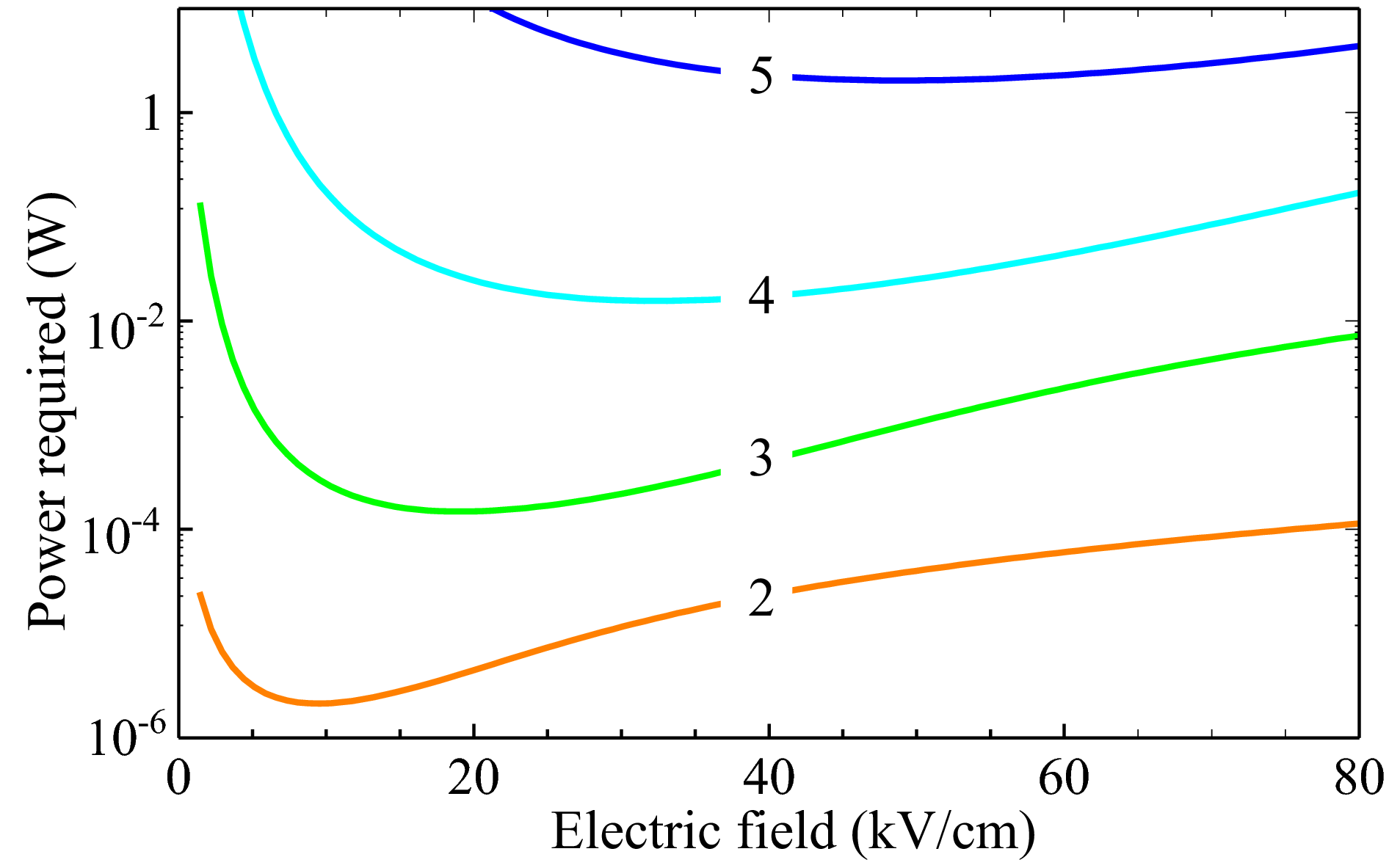}
\caption{
\label{Fig:RotationalTransitions}
Driving rotational transitions in a static electric field. Molecules enter the field in the $(N,M)=(0,0)$ state, and exit in the $(N,0)$ state. The power needed for a $\pi$-pulse is plotted as a function of the static electric field strength, for the experimental parameters given in the text. Curves are labelled by their value of $N$. Note the logarithmic scale.}
\end{center}
\end{figure}

We suppose that ground state YbF molecules, travelling at 340\,m/s, cross a resonant microwave beam at right angles. The microwaves have a cylindrical focus, 20\,mm long in the direction of the molecular beam, and 4\,mm in the perpendicular direction. Figure \ref{Fig:RotationalTransitions} shows the power required to drive a $\pi$-pulse to the state that correlates to the field-free $(N,M\!=\!0)$ state, as a function of the applied, uniform, static electric field strength\footnote{To apply this plot to a different rigid rotor molecule, $X$, multiply the power by $(\mu_{\rm{YbF}}/\mu_{X})^{2}$ and multiply the electric field by $(B_{X}/B_{\rm{YbF}})(\mu_{\rm{YbF}}/\mu_{X})$}. As anticipated by the considerations above, the required power is high at both low and high fields. The power requirement increases rapidly with $N$, and the electric field that minimizes this power also increases with $N$. For the $(5,0)$ state, the power is minimized in a static field of 49\,kV/cm, and is then 2\,W at 286\,GHz. This seems unfeasible. Furthermore, the power requirement increases if the transition is inhomogenously broadened, and it will be unless the static field is uniform to 1 part in $10^{7}$. A better route takes the molecule via the $(3,0)$ state. At 31\,kV/cm the transitions correlating to $(0,0)\rightarrow(3,0)$ and $(3,0)\rightarrow(5,0)$ have the same frequency, 126\,GHz. At this field strength, the power requirements are 232\,$\mu$W and 7\,$\mu$W respectively. Inhomogeneous broadening is severe for the first transition whose Stark shift is about 2GHz per kV/cm at this field. If the field in the interaction region is non-uniform at the $10^{-4}$ level, the power needed increases to about 200\,mW, a demanding, but not unfeasible, requirement.

\subsection{Alternating gradient deceleration in the ground state}\label{Sec:AG}

Whereas weak-field seeking molecules are naturally focussed through a Stark decelerator, strong-field seekers are naturally defocussed. The alternating gradient (AG) decelerator provides a solution to this focussing problem, as explained in detail in \cite{Bethlem(1)06} and \cite{Tarbutt(1)08}. The molecules move through an alternating sequence of focussing and defocussing lenses, always passing closer to the axis in the defocussing lenses than in the focussing lenses. Because the transverse force is proportional to the off-axis displacement, these molecules are on stable trajectories. In the ideal case, each lens focusses the molecules in one transverse direction, and defocusses them in the other, the force components being $F_{x}= k x$ and $F_{y}=-k y$. The two directions alternate in successive lenses. The stability of the trajectories, and the transverse phase space acceptance, are determined by the product $\Omega\tau$, where $\Omega=\sqrt{k/m}$ is the angular oscillation frequency inside a focussing lens, and $\tau$ is the time spent in each lens. When $\Omega\tau\ll 1$, the focussing by one lens is almost exactly counteracted in the next and the net effect is very weak focussing. The trajectories are stable but have large amplitude, so the acceptance is small. As $\Omega\tau$ increases, the transverse extent of the beam modulates, always being larger in the focussing lenses than in the defocussing lenses. Molecules with greater transverse speeds can now be transmitted, and so the acceptance grows. At some critical value of $\Omega \tau$, the focussing becomes so strong that the molecules are overfocussed and all trajectories become unstable.

\begin{figure}[t]
\begin{center}
\includegraphics[width=12cm]{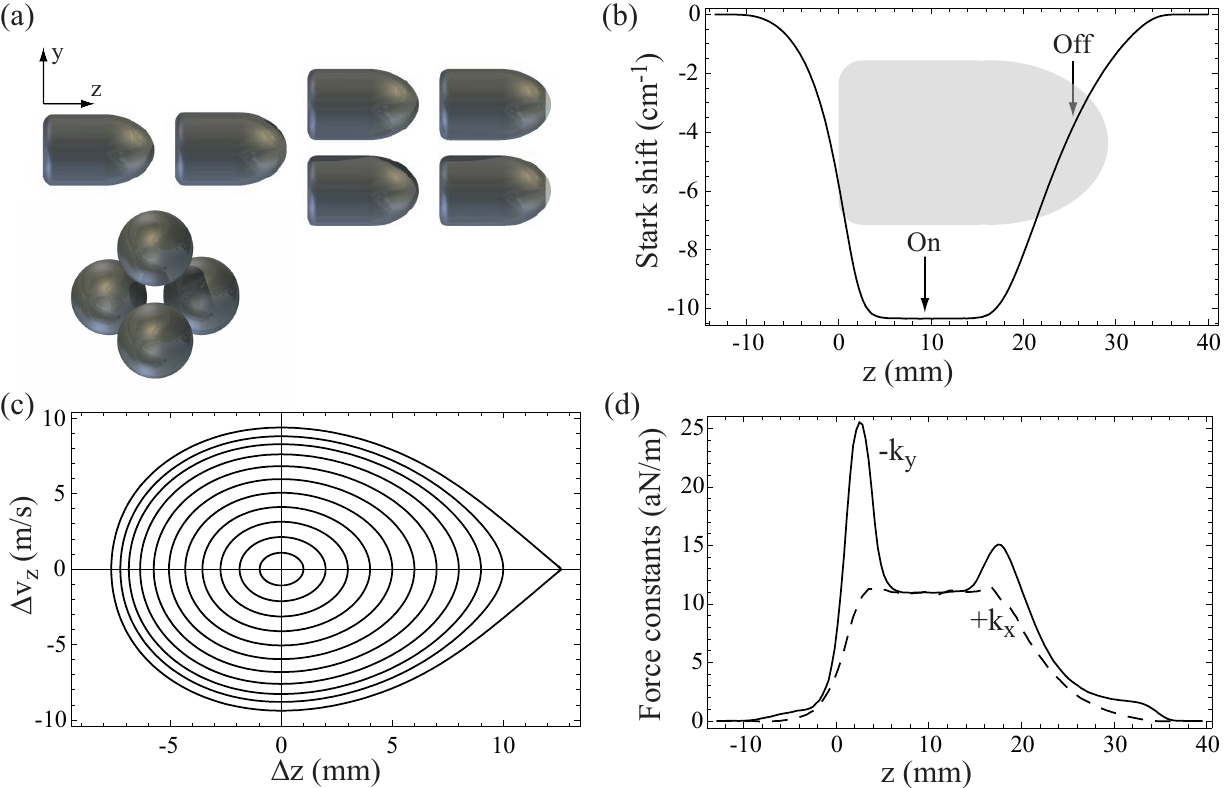}
\caption{
\label{Fig:AGPlots}
Alternating gradient deceleration of ground state YbF. (a) Projections of the decelerator onto the $y z$- and $x y$-planes. (b) The $z$-dependence of the Stark shift, with the position of the central electrode indicated in grey. The entrance end of this electrode is positioned at $z=0$. (c) Longitudinal phase-space acceptance calculated from the potential and the turn-on and -off positions shown in (b). (d) The focussing (dashed line) and defocussing (solid line) force constants, $k_{x}=-\partial F_{x}/\partial x$ and $k_{y}=-\partial F_{y}/\partial y$. For ease of comparison, $-k_{y}$ is plotted along with $+k_{x}$.}
\end{center}
\end{figure}

AG deceleration of ground state YbF has been demonstrated experimentally \cite{Tarbutt(1)04} using a short prototype machine. Other strong-field seeking molecules have also been decelerated using similar machines \cite{Bethlem(1)02,Wohlfart(1)08,Wohlfart(2)08}. In these experiments, the degree of deceleration has been limited and the transmission rather low. The AG decelerator is particularly sensitive to lens aberrations which tend to reduce its transmission enormously. Here, we consider a particular design where we aim to minimize the detrimental effects of aberrations and provide a very large reduction in the velocity. The design is illustrated in Fig.\,\ref{Fig:AGPlots}(a). Each lens is formed by a pair of rods, radius $R$, their axes parallel and separated by $2(R+r_{0})$, similar to the geometry used in all AG experiments to date. Specifically, we take $r_{0}=2$\,mm and $R=9.3$\,mm. At the exit end, each rod is terminated by a semi-ellipsoid of radii 9.3\,mm along $x$ and $y$, and 13.3\,mm along $z$. At the entrance, the edge of the rod is rounded with a 2.7\,mm radius of curvature. The total length of each electrode is 29.3\,mm, and the distance from the entrance of one lens to that of the next is 34.7\,mm.

The choice of $R/r_{0}$ is determined by a trade-off between the strength of focussing and the size of the non-linear contributions to the force \cite{Tarbutt(2)08}. When $R/r_{0}$ is large, the focussing will be weak and the acceptance will be small. When it is small the beneficial effect of the enhanced focussing is outweighed by the detrimental effect of large nonlinear forces. Our choice, $R/r_{0}=14/3$, is close to optimum for a decelerator where the length of the lenses is approximately equal to the length of the gaps between lenses, and where $\Omega\tau$ has its best value of about 0.7. As shown in Fig.\,11 of \cite{Tarbutt(2)08}, the transverse acceptance is then $0.012 \times 2 \mu_{\rm{eff}}E_{0}r_{0}^{2}/m$. For our parameters, this is 70\,mm$^{2}$\,(m/s)$^{2}$, a factor of 2 smaller than the transverse acceptance estimated for the WF decelerator above. A higher transverse acceptance can be achieved using a more complicated electrode structure to form each lens, but only by a factor of about 2.5 \cite{Tarbutt(2)08}.

The deceleration regions are the ellipsoidal ends of the rods. Figure \ref{Fig:AGPlots}(b) shows the $z$-dependence of the Stark shift for our lens structure, along with a cross-section of an electrode so that its position relative to the map is clear. The lens is charged to $\pm 40$\,kV, and its neighbours are grounded. To make a first estimate of the longitudinal acceptance, we consider only those molecules travelling along the axis of the machine, ignoring any coupling between transverse and longitudinal motions. Then, the position $\tilde z$ of a molecule relative to that of the synchronous molecule is given by the mean (relative) acceleration \cite{Tarbutt(1)08}

\begin{equation}
\frac{d^{2}\tilde z}{dt^{2}} = \frac{W(z_{{\rm on}} + \tilde z) -
W(z_{{\rm on}}) - W(z_{{\rm off}}+\tilde z) + W(z_{{\rm off}})}{m
D},
\end{equation}
where $W(z)$ is the $z$-dependent Stark shift shown in Fig.\,\ref{Fig:AGPlots}(b), $z_{{\rm on}}$ and $z_{{\rm off}}$ are the positions of the synchronous molecule when the fields of the lens are turned on and off, and $D$ is the lens-to-lens distance. By solving this equation of motion using the turn-on and turn-off positions indicated, we obtain the phase-space trajectories of non-synchronous molecules shown in Fig.\,\ref{Fig:AGPlots}(c). The outer trajectory is the separatrix between stable and unstable motion. Using this model, we obtain a longitudinal acceptance of 270\,mm\,m/s, almost 12 times as large as that for deceleration in the (5,0) state (Fig.\,\ref{Fig:WFAccept}). In the $\Delta z$ direction, the accepted region is about four times larger and is a direct result of the larger spacing between deceleration stages. In the $\Delta v_{z}$ direction, the accepted region is about three times larger, because of the larger Stark shift in the ground state. Within the model of uncoupled longitudinal and transverse motion, the total 6D acceptance of the AG decelerator is $18,900\,\rm{mm}^{3}(\rm{m/s})^{3}$.

In reality, the transverse and longitudinal motions are rather intimately coupled. We will arrange the lenses so that $\Omega \tau$ is close to the optimum value for the synchronous molecule. The non-synchronous molecules will have different values of this parameter. Referring to Fig.\,\ref{Fig:AGPlots}(b), a molecule that is ahead will spend less time in the lens and so will not be focussed as effectively, while a molecule that is behind will spend more time in the lens and may be overfocussed and become unstable. Furthermore, deceleration is always accompanied by transverse defocussing. This effect is shown in Fig.\,\ref{Fig:AGPlots}(d), where the transverse force constants $k_{x} = -\partial F_{x}/\partial x, \,k_{y} = -\partial F_{y}/\partial y$ are plotted as a function of $z$. Inside the lens, where there is no longitudinal electric field gradient, these two force constants are equal and opposite, $k_{x}=-k_{y}=11$\,aN/m. Moving into the fringe fields of the lens, the focussing force constant decreases along with the electric field, whereas the defocussing force constant increases to a value that depends on the longitudinal curvature of the electrodes. The smaller the radius of curvature, the larger this defocussing peak becomes. We have chosen a fairly gradual termination of the electrodes at the exit end, where the deceleration occurs, and the magnitude of $k_{y}$ peaks at 15\,aN/m. At the entrance end, the electrodes begin suddenly, and so the defocussing constant rises to a much larger value of 25\,aN/m. However, molecules that stay close to the synchronous phase never see this excess defocussing, because the voltages are not yet on when they pass through this region. Since the focussing is most effective when the gaps between successive lenses are kept short, it is good to keep this entrance region short; that is why we have opted for an asymmetry between the entrance and exit ends. When the turn-on and turn-off positions are as indicated in part (b) of Fig.\,\ref{Fig:AGPlots}, the mean value of $k_{y}$ seen by the synchronous molecule exceeds that of $k_{x}$ by 19\%. This makes the trajectories unstable if $\Omega \tau$ is too small, making it even more important to keep this parameter close to its optimum value. Provided that is done, the excess defocussing has a rather small effect. We note that molecules lagging behind the synchronous one see the very large defocussing constant at the lens entrance, and we can expect these to be lost from the decelerator.

\begin{figure}
\begin{center}
\includegraphics[width=12cm]{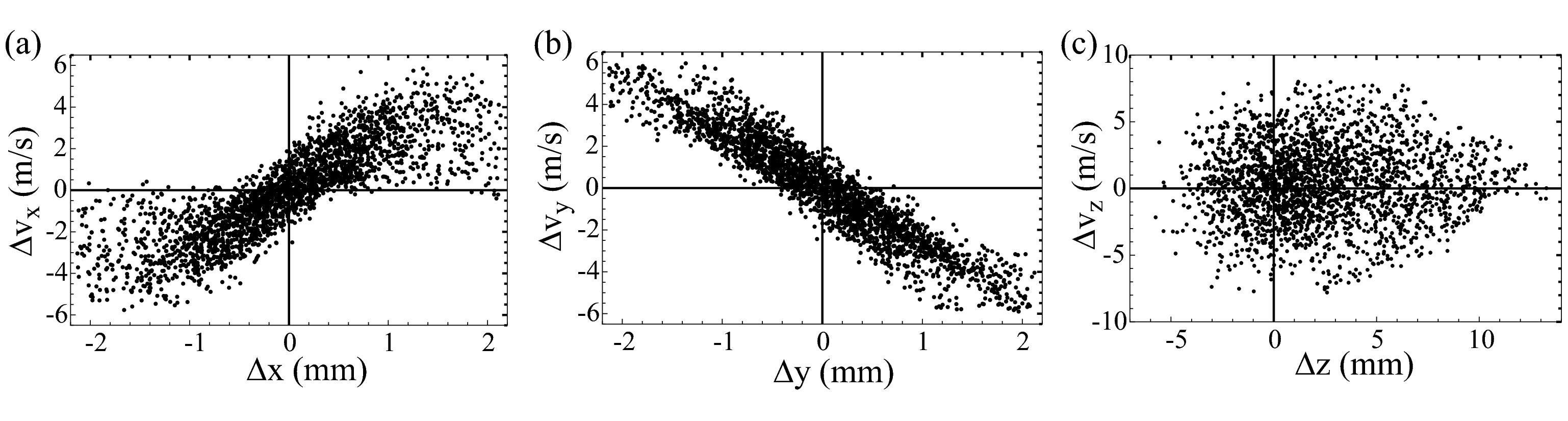}
\caption{
\label{Fig:AGAccept}
Phase-space acceptance of the first section of AG decelerator described in the text, projected onto the (a) $(x,v_{x})$, (b) $(y,v_{y})$ and (c) $(z,v_{z})$ planes. The section has 56 stages and reduces the speed from 340\,m/s to 266\,m/s.}
\end{center}
\end{figure}

If the turn-on and turn-off positions are kept constant, the value of $\tau$ increases as the molecules slow down. Eventually, the limit of stability is reached and the molecules are lost. To avoid that, the lens length seen by the molecules needs to be reduced in harmony with the deceleration: the lenses should be long at the beginning and short at the end. Since deceleration occurs at the end of every lens, it is better to use $n$ short focussing lenses, followed by $n$ short defocussing lenses, instead of using long lenses and alternating each time. By using a larger value of $n$ at the beginning than at the end, efficient deceleration can be combined with stability of focussing. We have simulated the motion of ground state YbF molecules through an AG decelerator having the electrode structure described above. The decelerator was divided into four sections, with $n=4,3,2,1$ and with $56, 36, 32, 15$ deceleration stages in these sections respectively. The first section reduces the speed from 340\,m/s to 266\,m/s, and the subsequent sections reduce this further to 205, 128 and 66\,m/s respectively. In order to avoid a mismatch between the phase-space distribution exiting one section and the acceptance of the next, the turn-on position was gradually adjusted within each section so that the value of $\Omega\tau$ remained approximately the same throughout the deceleration process.

Figure \ref{Fig:AGAccept} shows the accepted phase-space volume of the first section, projected onto the $(x,v_{x})$, $(y,v_{y})$ and $(z,v_{z})$ planes. The plots were obtained as described above for the WF decelerator. In the $(x,v_{x})$ plane the accepted molecules form a diverging beam because the first four lenses of the decelerator focus in this direction, while in $(y,v_{y})$ they form a converging beam because those lenses defocus in that direction. In the $(z,v_{z})$ plane, the shape of the accepted area is similar to the calculated one shown in Fig.\,\ref{Fig:AGPlots}(c), though a little smaller. In each plane, there is a dense central core of accepted molecules, with a less dense region near the edges. These regions of lower density are due to the coupling between the two transverse motions, and between the transverse and longitudinal motions. The acceptance of this first 56-stage section is $2750\,\rm{mm}^{3}(\rm{m/s})^{3}$, more than twice that found for the first 200 stages of the WF decelerator, where the final speed was nearly the same. Taken alone, the acceptances of the other three sections are similar, but when we concatenate them, the net acceptance falls. We calculate $850\,\rm{mm}^{3}(\rm{m/s})^{3}$ for the first two sections, $340\,\rm{mm}^{3}(\rm{m/s})^{3}$ for three sections and $254\,\rm{mm}^{3}(\rm{m/s})^{3}$ for all four. There are several causes for this additional loss. Firstly, those molecules near the edges of the acceptance region in Fig.\,\ref{Fig:AGAccept} are on metastable trajectories and will be lost in subsequent sections. Secondly, there remains some mismatch between one section and the next, despite our efforts to control this. Thirdly, as the turn-on position is moved towards the turn-off position the excess defocussing at the end of each lens plays an increasingly important role and can result in additional loss. Nevertheless, the acceptance of the AG design studied here is similar to that of the WF decelerator and it is likely that further design iterations will improve on this.

This 139-stage AG decelerator removes 96\% of the initial kinetic energy, but the molecules are still not slow enough to be trapped. Five more stages are needed. These final stages are particularly difficult to design because of the large fractional changes in speed, and they would need to be designed in conjunction with the trap to ensure that the molecules remain focussed and are coupled efficiently into the trapping region. An alternative is to use a combination of AG and WF deceleration. In this scheme, the AG decelerator would be used to remove the majority of the energy, and then the molecules would be switched into the weak-field seeking state for the final reduction of the velocity.

\section{Trapping}\label{sec:trapping}

The electric field of the trap is also the field that interacts with the edm being measured. The edm signal is proportional to the integral over time of the polarization factor, $P$, which in turn depends on the electric field as shown in Fig.\,\ref{Fig:Stark}(c). The trap should have a bias field so that $|P|$ remains large and approximately constant, i.e. close to its turning point for a weak-field seeking state, or close to unity for a strong-field seeking state.

It is not difficult to design a very deep, suitably-biased, electrostatic trap for weak-field seeking YbF molecules in the $(5,0)$ state.  As an example, we consider using the ``chain-link'' trap discussed in \cite{ShaferRay(1)03}. This has a minimum field at its centre, which we choose to be 130\,kV/cm in order to maximize the value of $|P|$ and minimize its variation. For modest applied voltages, the phase space acceptance of such a biased Stark trap is of order $10^{4}\,\rm{mm}^{3}(\rm{m/s})^{3}$, very much larger than the phase-space volume occupied by the molecules coming from the Stark decelerator. So, with a suitably careful coupling of the molecules into the trap, all the available slow molecules could be trapped and used in an edm measurement.

If we are instead to trap molecules in the strong-field seeking ground state, an ac trap is required. For this, we consider a cylindrical ac trap of the kind first proposed by Peik \cite{Peik(1)99} and recently used to trap ammonia molecules \cite{vanVeldhoven(1)05,Bethlem(2)06}. The centre of the trap is a saddle point, and the confining and deconfining directions alternate so that the molecules are trapped dynamically - the same principle that is used to guide molecules through the AG decelerator. An electric field of 100\,kV/cm at the centre of the trap would gives a polarization factor of 0.86. For the ammonia experiments, the acceptance of this trap was found to be $270\,\rm{mm}^{3}(\rm{m/s})^{3}$. Scaling to the YbF case, with its larger mass and larger dipole moment, we obtain an acceptance of about $50\,\rm{mm}^{3}(\rm{m/s})^{3}$. This small acceptance makes trapping in the ground state rather unattractive for an edm measurement. Additionally, the large currents associated with the switching of the high voltages are likely to be incompatible with the extremely high level of magnetic field control needed for the experiment.

\section{Statistical sensitivity}\label{sec:sensitivity}

We now have all the information needed to estimate the statistical sensitivity of a trap experiment relative to that of our beam experiment, using Eq.\,(\ref{Eq:SensitivityFormula}). It is convenient to write the total number of participating molecules as $N=\rho V r T$, where $\rho$ is the phase-space density, $V$ is the phase-space acceptance of the experiment, $r$ is the mean number of shots per second and $T$ is the total integration time. The first and last factors will be common to both beam and trap experiments and so factor out in the comparison. We introduce a figure of merit which is the ratio of the sensitivity in a trap experiment to that in our current beam experiment,

\begin{equation}
S=\frac{\delta d_{e,\rm beam}}{\delta d_{e,\rm trap}}=\frac{|P_{\rm trap}|}{|P_{\rm beam}|}\frac{\tau_{\rm trap}}{\tau_{\rm beam}}\sqrt{\frac{V_{\rm trap}}{V_{\rm beam}}}\sqrt{\frac{r_{\rm trap}}{r_{\rm beam}}}.
\end{equation}
In the beam, the volume of phase space occupied by the participating molecules is $V_{\rm beam}\approx 6\times 10^{4}\,\rm{mm}^{3}(\rm{m/s})^{3}$. The other parameters are $r_{\rm beam}=12.5\,\rm{s}^{-1}$, $|P_{\rm beam}|=0.7$ and $\tau_{\rm beam}=10^{-3}$\,s.

Consider first an experiment where the number of available molecules is limited by the acceptance of the WF Stark decelerator modelled above, which we found to be $350\,\rm{mm}^{3}(\rm{m/s})^{3}$. We take a coherence time in the trap of 1\,s, and we suppose that the trap is reloaded every 1.2\,s. With a polarization factor of 0.36 we obtain a figure of merit of $S\approx 10$. This would be a very significant improvement in the sensitivity of the experiment, offering the possibility of reaching well below $10^{-28}$e.cm in statistical uncertainty. Note that the repetition rate of the proposed trap experiment is limited by the coherence time, not the pulse rate of the source, and so its sensitivity could be improved even further by loading multiple traps, though at a cost of further experimental complexity.

Finally, let us consider an experiment that is limited by the acceptance of the trap itself rather than the method used to fill the trap. This may become possible by loading the trap from a buffer gas source, or from an improved decelerator design. In this case $V_{\rm trap}\approx 10^{4}\,\rm{mm}^{3}(\rm{m/s})^{3}$ and we find $S\approx 50$. All other factors being equal, in particular the phase space density produced at the source, the statistical sensitivity would then be about $2\times 10^{-29}\,e$\,cm after $T=24$\,hours of integration. To realize such exceptional sensitivity we would, of course, need to reduce all other sources of noise to this level. We see no fundamental reason why that should not be possible.


\section{Physics in the trap}

So far we have considered how the statistical sensitivity of an edm experiment might be improved by decelerating YbF molecules and loading them into a trap. Now we consider the impact of the trap environment on such an experiment. Our present edm experiment measures the precession of the molecular spin in a combination of nominally parallel electric and magnetic fields. We prepare the spin in a direction perpendicular to the fields and, after a fixed time, we measure the accumulated precession angle, which depends on the magnetic moment and the electron edm. In the trap, the electric field is necessarily inhomogeneous - that is how it exerts a force. Therefore, as molecules move around within the trap, they experience a substantial variation in the magnitude and direction of the field and this fluctuating environment varies from one molecule to another. It is therefore essential to consider what happens to the spin precession angle when the direction of the local electric field at the position of the molecule is changing with time. In the following, we consider two consequences of this: an effective magnetic field inhomogeneity, and a geometric phase.

\subsection{Effective magnetic field inhomogeneity}

Because of the large tensor Stark shift induced by the electric field of the experiment, only the magnetic field component parallel to the local electric field direction contributes to the precession angle \cite{Hudson(1)02}. The accumulated phase angle due to a magnetic field $\vec{\bf{B}}$ is
\begin{equation}
\phi_{B}=\frac{g \mu_{B}}{\hbar}\int \vec{\bf{B}}(\vec{\bf{r}}(t))\cdot\vec{\bf{\varepsilon}}(\vec{\bf{r}}(t))\,dt,
\label{Eq:MagneticPhase}
\end{equation}
where $g$ and $\mu_{B}$ are the g-factor and Bohr magneton, $\vec{\bf{\varepsilon}}$ is a unit vector along the local electric field direction, and the integral is taken over the trajectory of the molecule, $\vec{\bf{r}}(t)$.

In a biased electric trap (Sec.\,\ref{sec:trapping}), the trapping field is superimposed on a uniform bias electric field. We define $\hat{\bf{z}}=\vec{\bf{\varepsilon}}(0)$, the electric field direction at the centre of the trap. For those molecules that stay close to the centre, the local electric field direction is always close to $\hat{\bf{z}}$, whereas molecules that reach the outer regions of the trap will see $\vec{\bf{\varepsilon}}$ making larger angles with $\hat{\bf{z}}$. It follows that the integral in Eq.\,(\ref{Eq:MagneticPhase}) changes from one trajectory to the next {\it even if the magnetic field is perfectly uniform}. This spread in the accumulated phase is a source of noise in the experiment and may degrade the experimental sensitivity if it is too large. The size of the phase spread depends on the temperature of the molecules relative to the Stark shift at trap centre (expressed in temperature units). If the temperature is high the molecules will explore a large range of electric field directions and the spread will be high. If cold, the molecules remain close to the trap centre where the electric field direction is close to $\hat{\bf{z}}$ everywhere.

To explore this, we simulated the motion of molecules in the chain-link trap discussed in Sec.\,\ref{sec:trapping}, and evaluated $\phi_{B}$ for each trajectory using Eq.\,(\ref{Eq:MagneticPhase}). The molecules were drawn from an initial phase space volume of $(4\,\rm{mm}\times 4\,\rm{m/s})^{3}$. As discussed above, the molecules are most sensitive to fields along $\hat{\bf{z}}$ and so we applied a uniform magnetic field, $B_{z}$, in this direction. We find that both the mean and the standard deviation of the accumulated phases increase linearly with the coherence time, $\tau$. Specifically we find a mean of $\langle \phi_{B}\rangle\approx0.9\,\phi_{0}$ and a standard deviation of $\Delta\phi_{B}\approx 0.05\,\phi_{0}$, where $\phi_{0}=\,g \mu_{B} B_{z} \tau/\hbar$ is the phase accumulated by a molecule that remains at the centre of the trap throughout. This phase spread should be compared to the error in a shot-noise limited measurement of the edm phase, which is $1/\sqrt{N}$\,rad (see Eq.\,(\ref{Eq:SensitivityFormula})). Therefore, to ensure that the magnetic field does not increase the noise and so degrade the sensitivity, we require $\Delta\phi_{B}<1$. For $\tau=1$\,s, the corresponding requirement on magnetic field is $B_{z} < 200$\,pT. Magnetic field gradients will also contribute noise as different molecules sample different regions of the trap. Our simulations show that, for a coherence time of 1\,s, these gradients need to be smaller than 50\,nT/m if they are not to degrade the sensitivity. These field requirements are not too demanding; they can be satisfied using a multi-layer magnetic shield with an overall shielding factor of $10^{6}$, e.g. \cite{Budker(1)98}.

\subsection{Geometric phase}

We show below that, for a sufficiently strong electric field, the motion of the spin follows the electric field direction adiabatically, and that the spin polarization will always lie in the plane perpendicular to the local electric field. In this adiabatic limit, one might guess that the accumulated phase of the spin would be the time-integral of the instantaneous precession frequency. However, in the case of a rotating magnetic field, it is known from the work of Berry \cite{Berry} that the accumulated phase angle has a second component, known as the geometric phase. This depends on the history of the field direction but not on the precession dynamics. We might expect that a similar geometric phase will appear in the spin precession of a molecule when the electric field changes direction. If we are to obtain the electron edm from the precession angle, we must be able to control any such geometric phase with high precision.

While the geometric phase of a spin precessing in a rotating magnetic field has been studied in great detail \cite{Berry}, the case of a rotating electric field is less well-studied and the few treatments that do exist tend to be phrased in the somewhat arcane language of differential geometry \cite{Wilczek}. In the interest of clarity, therefore, we first present a simple quantum mechanical treatment of the geometric phase in a rotating electric field, before going on to consider the impact of this phase on a potential edm experiment.

In the limit of large electric field, which is well-fulfilled in our experiments, the magnetic field makes no significant contribution to the geometric phase evolution. This is because the large tensor Stark shift strongly suppresses the interaction with magnetic field components perpendicular to the electric field \cite{Hudson(1)02}. Therefore we neglect the applied magnetic field in the following treatment, even though it is possible in principle to include it \cite{Hoodbhoy}.

\subsubsection{Spin-one system in a rotating electric field}
\label{geometric_phase}

Let us consider a spin-one system, corresponding to the $F=1$ state of the YbF molecule used in our edm experiment. The Stark interaction mixes many states of the molecule, but for our present purpose it is adequately described by an effective Hamiltonian restricted to the $F=1$ manifold, consisting of a scalar and a rank-2 tensor. With the electric field directed along the z-axis this takes the form
\begin{equation}
\hat{H}_{E_z} = \delta(E)\,\hat{T}^{(0)}_0 + \epsilon(E)\,\hat{T}^{(2)}_0\ ,
\end{equation}
where the coefficients $\delta(E)$ and $\epsilon(E)$ are phenomenological parameters that depend on the electric field strength $E$. If desired, they can be derived from a full calculation of the Stark shift including all of the molecular levels. $\hat{T}^{(n)}_0$ is the z-component of the rank-$n$ irreducible spherical tensor operator. The matrix elements of this operator in the $\ket{F=1, m_F}$ basis can be written, in order of ascending $m_F$, as
\begin{equation}
\hat{H}_{E_z} = \begin{pmatrix}0 & 0 & 0 \\ 0 & \Delta & 0 \\ 0 & 0 & 0 \end{pmatrix}\ ,
\end{equation}
where we have simplified the parametrization by redefining the zero of energy and introducing $\Delta = \sqrt{1/3}\,\delta(E) -\sqrt{2/15}\,\epsilon(E)$. We have dropped the explicit functional dependence of the parameter $\Delta$ on the electric field in order to streamline the notation.

When the electric field rotates to a new direction given by the Euler angles $\{ \alpha(t), \beta(t), \gamma(t)\}$, the Hamiltonian becomes
\begin{equation}
\hat{H}(t) = R(t)^{-1}\,\hat{H}_{E_z}\,R(t)\ ,
\end{equation}
where $R(t)$ is the rotation operator. We adopt the Euler angle convention of Weissbluth \cite{Weissbluth}: a rotation about the $z$-axis by $\gamma$, a rotation about the $y'$-axis by $\beta$, and finally a rotation about the $z''$-axis by $\alpha$. The rotation matrix acting on states and operators (rather than on coordinates) is then given by
\begin{equation}
\label{rot_mat}
R(t) =
\left(
\begin{array}{lll}
 e^{i (\alpha +\gamma )} \cos ^2\left(\frac{\beta }{2}\right)
 & \frac{1}{\sqrt{2}} e^{i \alpha } \sin \beta
 & e^{i (\alpha -\gamma )}\sin ^2\left(\frac{\beta }{2}\right) \\
 -\frac {1}{\sqrt{2}}e^{i\gamma} \sin \beta
   & \cos \beta
   & \frac {1}{\sqrt{2}}e^{-i\gamma} \sin \beta \\
 e^{i (\gamma -\alpha )} \sin ^2\left(\frac{\beta }{2}\right)
 & -\frac {1}{\sqrt{2}}e^{-i\alpha} \sin \beta
 & e^{- i (\alpha + \gamma )}
   \cos ^2\left(\frac{\beta }{2}\right)
\end{array}
\right)\ .
\end{equation}
We note that the cylindrical symmetry of $\hat{H}_{E_z}$ around $z$ also makes $\hat{H}(t)$ symmetric around $z''$ and therefore independent of $\alpha$.\footnote{Explicitly,
\begin{equation}
\hat{H}(t) =\frac{1}{2\sqrt{2}}
\left(
\begin{array}{lll}\nonumber
 \sqrt{2} \Delta  \sin ^2\beta
 & -e^{-i \gamma} \Delta \sin 2\beta
 & -\sqrt{2} e^{-2 i \gamma} \Delta  \sin ^2\beta \\
  -e^{i \gamma} \Delta \sin 2\beta
 & 2\sqrt{2}\Delta  \cos^2\beta
 & e^{-i \gamma} \Delta \sin 2\beta\\
 -\sqrt{2} e^{2 i \gamma} \Delta  \sin ^2\beta
 & e^{i \gamma} \Delta \sin 2\beta
 & \sqrt{2} \Delta  \sin ^2\beta
\end{array}
\right)\,.\
\end{equation}}
The Schr\"odinger equation takes the form
\begin{equation}
\label{Eq:labSchrodinger}
R(t)^{-1}\hat{H}_{E_z}R(t)C(t)=i\hbar \frac{\partial}{\partial t}C(t)\,,
\end{equation}
where $C(t)$ is the column matrix of coefficients $c_{m_F}$ from the expansion $\ket{\psi} = \displaystyle\sum_{m_F} c_{m_F}(t) \ket{{m_F}}$. The same state $\ket{\psi}$ can be expanded on the basis that rotates with the electric field and has $z''$ as its quantization axis. This has expansion coefficients  $C_R(t)=R(t)C(t)$. In terms of these, Eq.~(\ref{Eq:labSchrodinger}) becomes
\begin{equation}
R(t)^{-1}\hat{H}_{E_z}R(t)R(t)^{-1}C_R(t)=i\hbar\frac{\partial}{\partial t}R(t)^{-1}C_R(t)\,,
\end{equation}
which simplifies to
\begin{equation}
\left(\hat{H}_{E_z}+\hbar\,\hat{G}(t)\right)C_R(t)=i\hbar\frac{\partial}{\partial t}C_R(t)\,,
\end{equation}
where
\begin{equation}
\label{g_op}
\hat{G}(t) = -iR(t)\frac{\partial}{\partial t}R(t)^{-1}\,.
\end{equation}
We see that the evolution of the state expressed in the rotating $z''$ basis is governed by an effective Hamiltonian made up of the static $\hat{H}_{E_z}$, plus a term $\hbar\hat{G}(t)$ involving the time-dependence of the rotation. This second term is much like a fictitious force in classical mechanics. As we shall see, it generates a geometric phase.

We can write $\hat{G}(t)$ explicitly, using (\ref{g_op}) and (\ref{rot_mat}):
\begin{equation}
\hat{G}(t) = \frac{1}{\sqrt{2}}\left(
\begin{array}{lll}
 -\sqrt{2}(\dot{\alpha}+\dot{\gamma}\cos \beta)
 & e^{i \alpha} (\dot{\gamma}\sin\beta+i\dot{\beta})
   & 0 \\
 e^{-i \alpha} (\dot{\gamma}\sin \beta -i\dot{\beta})
 & 0
   & e^{i \alpha}(\dot{\gamma}\sin \beta +i \dot{\beta})\\
 0 &  e^{-i \alpha}(\dot{\gamma}\sin \beta -i \dot{\beta})
   & \sqrt{2}(\dot{\alpha}+\dot{\gamma}\cos \beta)
\end{array}
\right)\ ,
\end{equation}
where dotted quantities are derivatives with respect to time. We wish to consider the adiabatic limit, where the tensor Stark splitting $\Delta$ is always much larger than the rotation rate. In this limit, it is a good approximation to replace $\hat{G}(t)$ by
\begin{equation}
\label{g_mat}
\hat{G}_a(t) = \left(
\begin{array}{ccc}
 -\dot{\alpha}-\dot{\gamma}\cos \beta & 0 & 0 \\
 0 & 0 & 0 \\
 0 & 0 & \dot{\alpha}+\dot{\gamma}\cos \beta
\end{array}
\right)\,.
\end{equation}
The subscript $a$ indicates that this is the effective operator in the adiabatic limit.
This approximation follows from the formal properties of matrices that the eigenvalues and eigenvectors of $\bigl( \begin{smallmatrix}a & b & 0 \\ b^* & \Delta & b \\ 0 & b^* & -a \end{smallmatrix} \bigr)$ approach those of the diagonal matrix $\bigl( \begin{smallmatrix}a & 0 & 0 \\ 0 & \Delta & 0 \\ 0 & 0 & -a \end{smallmatrix} \bigr)$ in the limit of large $\Delta$. It is also familiar from time-independent perturbation theory, which applies in the limit of slow rotations, where the effects of the off-diagonal elements $G_{ij}$ are smaller than those of the diagonal elements $G_{ii}$ by a factor of order $G_{ij}/\Delta$.

The operator $\hat{G}_a(t)$ is the central result of this section: in the adiabatic limit, the dynamics of a spin-one system in a rotating electric field are the same as those in the static field plus a time-dependent fictitious magnetic field directed along the electric field. This breaking of degeneracy between the $m_F = \pm 1$ sub-levels has nothing to do with the static interaction. On the contrary, it is a vector property caused by the vector nature of rotation\footnote{Rotation is generated by angular momentum, a vector quantity.}. We also note that the adiabatic condition only requires the splitting $\Delta$ between the $m_F=0$ and $m_F=\pm 1$ sub-levels to be large compared with the rotation rates. The splitting between $m_F = \pm 1$ sub-levels is unimportant as the angular momentum operator does not couple these states.

In order to calculate $\hat{G}_a$ from Eq.~(\ref{g_mat}), we need the Euler angles as a function of time.  The electric field rotation fixes $\beta(t)$ and $\gamma(t)$ unambiguously, but we have complete freedom to choose $\alpha(t)$ because of the cylindrical symmetry of the Stark interaction around $z''$. It is convenient to pick $\alpha(t)$ such that $\hat{G}_a$ vanishes, which requires
\begin{equation}
\label{connection}
\alpha (t) = \alpha_0 - \int_0^t \dot{\gamma}(\tau)\cos (\beta (\tau))  \,d\tau\ ,
\end{equation}
where $\alpha_0$ is a constant of integration. In this particular rotated basis, the adiabatic Hamiltonian is just that of the molecule in a static electric field. However, when the field makes some excursion before returning to its original direction, the final value of $\alpha$, given by (\ref{connection}), will not be the same as the starting value $\alpha_0$. This is the essence of the geometric phase.

\begin{figure}[tb]
\center
\includegraphics[width=6cm]{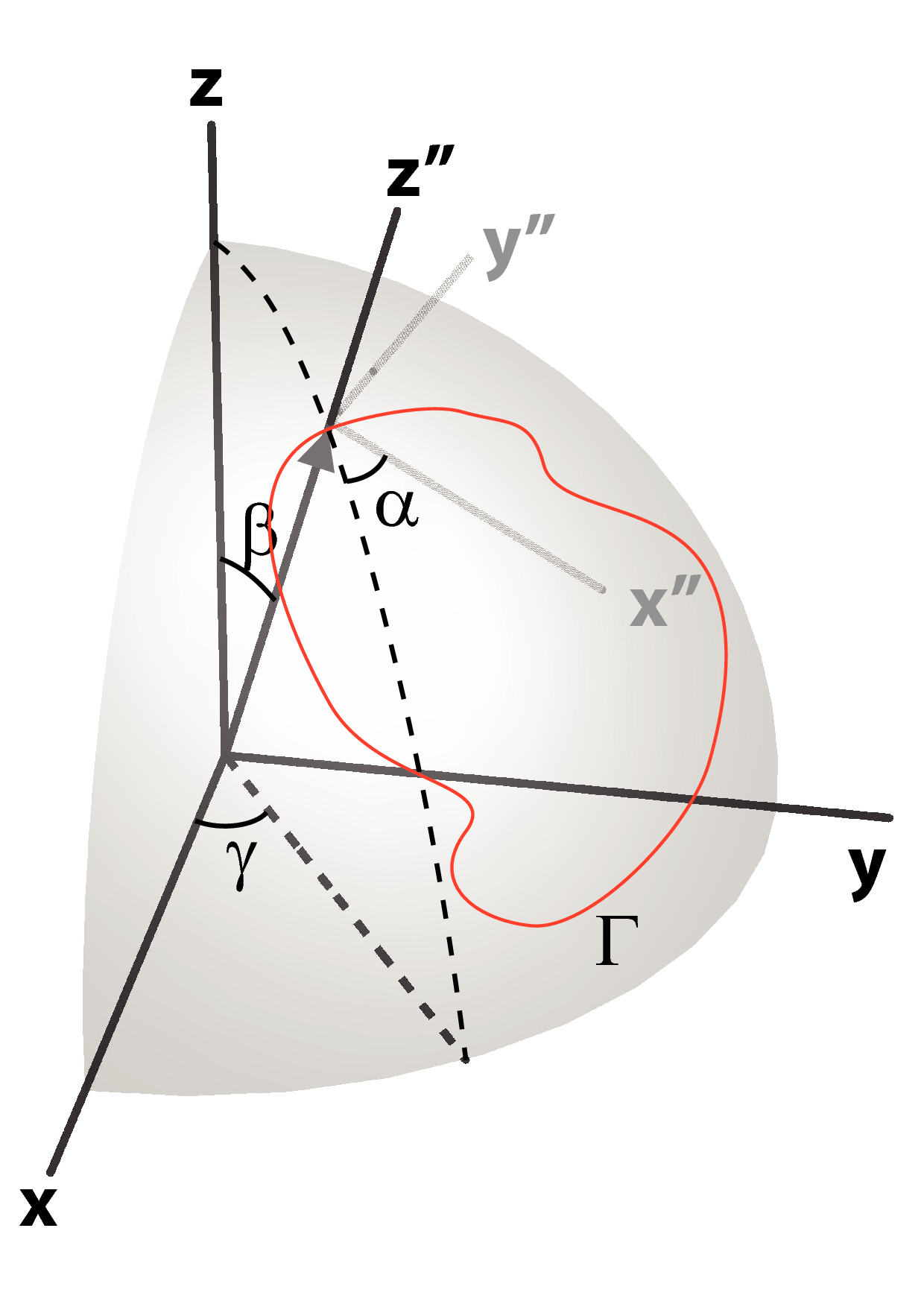}
\caption{\label{tangent_plane}
Geometry of the problem. The $z''$ axis is along the electric field direction defined by $(\beta,\gamma)$. When the field rotates, a unit vector along $z''$ follows the path $\Gamma$ (red line) on the surface of a unit sphere.  The $x''$ axis is tangent to the sphere and lies at an angle $\alpha$ to the dashed great circle. This angle changes under parallel transport around the loop, as given by Eq.~(\ref{connection}), and does not in general return to its original value $\alpha_0$ at the end of the loop. The change in $\alpha$ is equal to the solid angle subtended at the origin by the red curve $\Gamma$.}
\end{figure}

Figure \ref{tangent_plane} illustrates the geometry of our problem. A unit vector along $z''$ follows the rotation of the electric field, mapping out the path $\Gamma$ on the surface of a sphere. At any given point on the path, with coordinates $(\beta,\gamma)$, the $x''$ axis lies in the tangent plane, and makes an angle $\alpha$ to the great circle of constant $\gamma$. Following an initial choice of $\alpha_0$, the subsequent evolution of $\alpha$ under Eq.~(\ref{connection}) is given by \emph{parallel-transport} of the $x''$-$y''$ plane along the path $\Gamma$. This means that for each infinitesimal displacement along the path, the new basis is the one most nearly parallel to the old basis. In the interest of brevity, we simply state this here and do not prove it \cite{proof}. At the end of a rotation around a closed path $\Gamma$, it is easy to verify that $\alpha-\alpha_0$, as given by Eq.~(\ref{connection}), is equal to the solid angle $\Omega$ subtended at the origin by the curve\footnote{More correctly, it is $\Omega-2\pi$, but the $2\pi$ comes from our definition of the Euler angles and has no physical consequence}. This is an example of a more general result from differential geometry, that the so-called holonomy angle introduced by parallel-transport around a curve is equal to the integral of the Gaussian curvature over the region bounded by the curve. For a sphere, which has constant Gaussian curvature $1/r^2$, this integral is equal to the solid angle.

This phase shift has a mechanical consequence. Consider a spin prepared at $t=0$ along $x=x''(0)$. Under adiabatic rotation of the electric field, the spin undergoes parallel transport into the new $x''$ direction. At the end of a closed loop, although the spin has returned to the initial $x-y$ plane, its direction $x''(t)$ is now rotated through an angle $\alpha(t)-\alpha_0=\Omega$, equal to the solid angle subtended by the path $\Gamma$. It is this spin rotation that poses the main challenge for measuring the electron edm using trapped YbF.

\subsubsection{Geometric phase in the trap}

\begin{figure}[tb]
\center
\includegraphics[width=12cm]{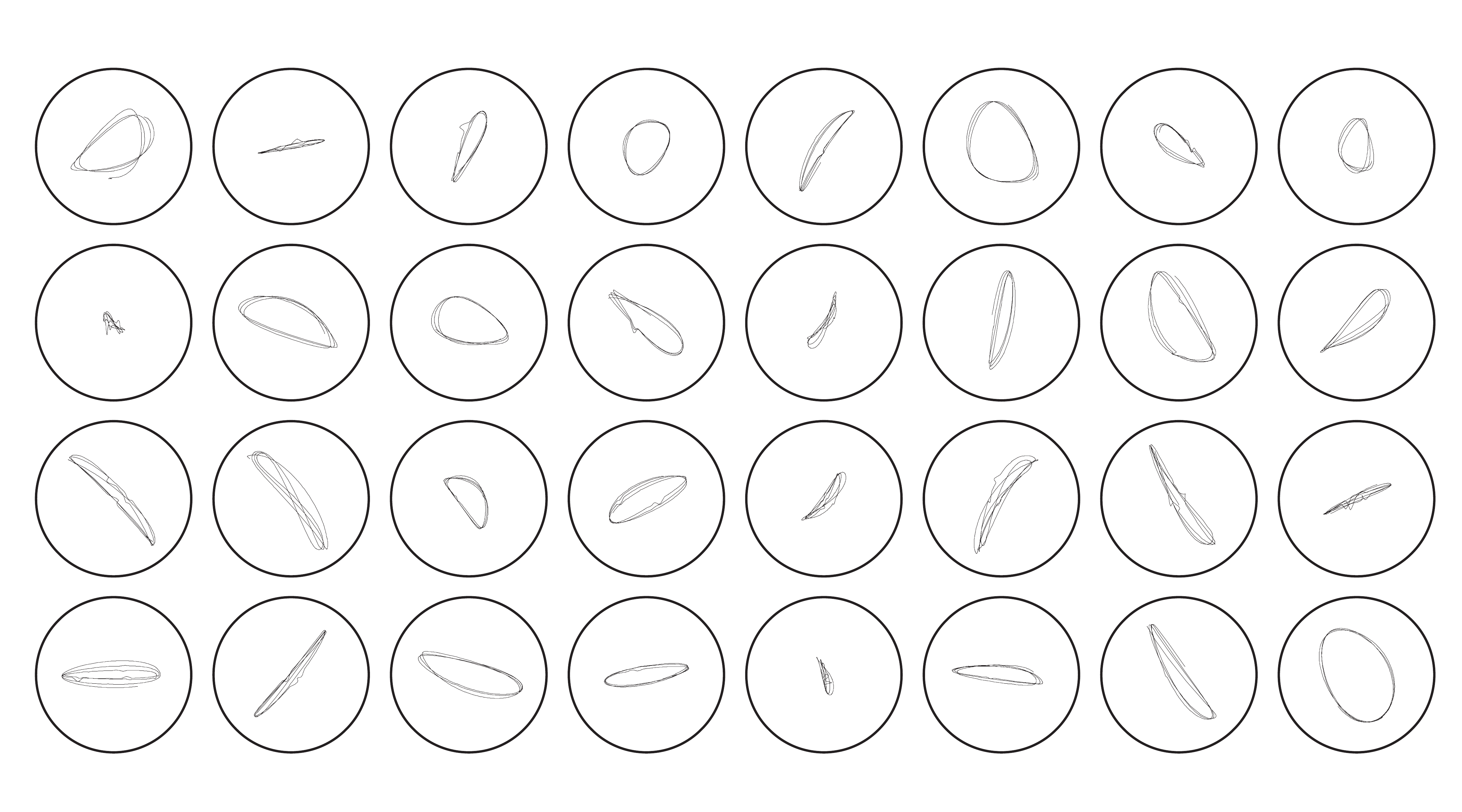}
\caption{\label{e normed}
$x$-$y$ projection of the normalised local electric field at the positions of 32 molecules evolving in the trap for 10\,ms. Each circle shows the curve swept out on the sphere by the tip of the electric field direction vector for a single molecule, as viewed from the north pole.}
\end{figure}

We have made a numerical simulation of thirty-two molecular trajectories in the chain-link trap of Sec.~\ref{sec:trapping} in order to evaluate the effect of the geometric phase. This calculation was made as realistic as possible by using finite-element-analysis to determine the field. The molecules were loaded into the trap with a uniform distribution over $\pm 2$\,mm in all three spatial directions and over $\pm 2$\,m/s in all three velocity components. Figure \ref{e normed} shows $x$-$y$ projections of the unit vector along the electric field at the position of each molecule. These views, seen along the the $z$-axis, show the path $\Gamma$ (illustrated schematically in Fig.~(\ref{tangent_plane})) that determines the geometric phase. It is immediately clear without further calculation that the accumulated geometric phases are large and different for each molecule. The solid angle for each orbit of the trap can approach a sizeable fraction of $2\pi$. These simulations last for 10\,ms, during which time the molecules execute 3 or 4 oscillations. The result is a large inhomogeneous spread of the spin direction in the $x$-$y$ plane that completely depolarizes the molecular ensemble. Therefore the spin decoherence time is only a few ms, not the 1\,s proposed in Sec.~\ref{sec:sensitivity}. The geometric phase is clearly the primary obstacle to making an edm measurement, or indeed any other spin precession measurement, in an electrostatic trap.

\section{Conclusions and outlook}

The use of trapped polar molecules to measure the electron edm offers a very substantial increase in statistical sensitivity. By loading an electrostatic trap using a Stark decelerator a sensitivity gain of 10 is feasible. Deceleration is possible in either the ground state or a rotationally excited state with comparable efficiencies. The efficiency of AG deceleration is particularly sensitive to the electrode geometry and timing sequence and it seems likely that its losses could be reduced further through refinement of these parameters. A buffer gas source might be able to deliver more molecules to the trap and so increase the statistical sensitivity even further.

However, the trap is a difficult environment for an edm experiment. The spatial variation in the direction of the electric field which provides the trapping force has two deleterious effects. The first is an inhomogeneous broadening of the magnetically-induced spin-precession angle, which can be kept under control by making the applied magnetic field small enough. The second is an inhomogeneous geometric phase. The long coherence time that seemed to be possible in the trap is precluded by this geometric phase which scrambles the spin precession signal in just a few milliseconds.

One possibility for overcoming this problem might be to use a molecule with $\Omega$-doublet structure, for example ThO \cite{vut08a}. In these molecules there are states with equal $m_F$ but opposite sensitivity to the edm. A superposition of two such states prepared within the same molecule would experience a common geometric phase shift but a differential edm phase shift. This might provide the basis for a reduced sensitivity to the geometric phase, but it would be vulnerable to differential Stark shifts between the two $\Omega$-doublet levels resulting from the variation in electric field strength. Further analysis is required to determine if this could be used as the basis for a workable edm measurement. A molecule with a pure spin 1/2 ground state, and therefore no tensor Stark splitting, would not adiabatically follow the electric field and so would be immune to the geometric phase arising from the rotating electric field. However, the molecule would lose its anisotropic response to magnetic fields which is a key feature for suppressing both the magnetic geometric phase and the troublesome $v\times E$ effect \cite{Regan(1)02}. Another possibility is to use the decelerator as a source for a molecular fountain where it would seem to be much easier to control the electric and magnetic fields and thereby control the systematic effects. The coherence time approaches 1\,s in a 1\,m-high fountain. Note that the molecular beam exiting the decelerator would need to be expanded to $\approx$10\,cm diameter to obtain the required degree of collimation.

The difficulties associated with the trap environment occur because the temperature of the molecules is similar to the trap depth and so they explore most of the trap. If the molecules were far colder they would remain near the trap centre where the field is very uniform. This might be achieved by loading from a source with much higher phase-space density, so that we could afford to discard all but the coldest fraction, or by actively cooling the molecules to ultralow temperatures. There has been rapid progress in the formation of deeply-bound ultracold polar molecules using photoassociation, Feschbach resonance, and coherent transfer techniques \cite{Sage(1)05,Ospelkaus(1)08,Deiglmayr(1)08}, and these methods might be extended to molecules suitable for edm measurement. Several research groups are now exploring the direct application of laser cooling techniques to molecules \cite{DiRosa}. In this context we note that the vibrational Franck-Condon structure of YbF \cite{Linton} is favourable for cycling transitions. Other routes to ultracold polar molecules include sympathetic cooling \cite{Lara(1)06} and cavity-assisted cooling \cite{Domokos(1)02}. If a scheme can be devised to cool a suitable heavy polar molecule to low temperature, edm measurements would not just be improved but revolutionized.
\\

\noindent{\bf Acknowledgements.} We are grateful to the Royal Society and to the STFC for supporting us.

\noindent{\bf Note added:} we thank Prof.\ Shafer-Ray for bringing to our attention his independent, closely-related work on the ``Effect of the geometric phase on the possible measurement of the electron's electric dipole moment using molecules confined by a Stark gravitational trap'' \cite{NSR}.


\begin{thebibliography}{25}
\bibitem{proof} This result can be proved using elementary geometry and calculus. At each point on the curve $\Gamma$, one needs to consider the cone whose apex lies on the $z$ axis and whose surface is tangent to the sphere at that point. One then uses the facts that the cone is developable, and that lines of constant $\gamma$ are great circles.
\end{thebibliography}
\end{document}